\documentclass[12pt]{article}
\usepackage{graphics}
\usepackage{epsfig}

\begin{document}

\renewcommand{\thefootnote}{\fnsymbol{footnote}}

\begin{titlepage}

\begin{flushright} 
IC/2001/24 \\
DFAQ-2001/TH-03 \\
HIP-2001-14/TH \\
April 2001 
\end{flushright} 
\vspace{2cm}

\begin{center}

{\Large \bf TeV scale unification in four dimensions \\  
versus extra dimensions }

\vspace{1cm}
{\bf Zurab Berezhiani,$^{a,d,}$\footnote{berezhiani@fe.infn.it, 
berezhiani@aquila.infn.it }  
Ilia Gogoladze $^{b,d,}$\footnote{iliag@ictp.trieste.it} 
\  and  \  
Archil Kobakhidze $^{c,d,}$\footnote{Archil.Kobakhidze@helsinki.fi }} 
\vspace{1.5cm}

{$^a$ \it Universit\'a di L'Aquila, I-67010 Coppito, AQ and \\
 INFN, Laboratori Nazionali del Gran Sasso, I-67010 Assergi AQ, Italy \\ 
$^b$ International Centre for Theoretical Physics, Trieste 34100, Italy\\
$^b$ HEP Division, Department of Physics, University of Helsinki and\\
Helsinki Institute of Physics, FIN-00014 Helsinki, Finland\\
$^d$ Andronikashvili Institute of Physics, Georgian Academy of Sciences, \\
380077 Tbilisi, Georgia~} \\

\vspace{3cm}
\end {center}
\begin{abstract}
The gauge coupling constant unification at the low ($O$(TeV)) 
scale can be obtained just in four dimensions, without help 
of the power like renormalization group evolution in extra 
dimensions, due to the presence of some extra particle states 
at intermediate scales. 
We show explicit examples of such extra states  
in the range of 100 GeV -- 1 TeV which can be easily observed 
on future colliders and can have important impact on 
the particle phenomenology.   
The problems of the low scale grand unification and proton 
stability are also discussed.  
\end{abstract}

\end{titlepage}

\renewcommand{\thefootnote}{\arabic{footnote}}

\section{Introduction}

The gauge hierarchy problem is one of the most challenging problem in
particle physics. It questions why the electroweak scale 
$M_{W}\sim 100$ GeV is so small as compared 
to the scale of gravitational interaction $M_{P}\sim 10^{18}$ GeV 
which is often considered as a fundamental scale of physics.
Recently, however, it was shown in ref. \cite{ADD,AADD} 
that the essence of the hierarchy problem 
can be drastically changed in the presence of
reasonably large extra dimensions. In particular, in the context of $D$ 
-dimensional theory, with 1 time and $D-1$ spatial coordinates, one can
consider a picture when the Standard Model (SM) particles are localized on a
3-brane identified with the observed 3-dimensional space, while gravity
propagates in the full $D$-dimensional bulk with $N=D-4$ compact dimensions.
In this situation the fundamental scale of gravitational interaction could
be as low as $M_{Pf}\sim $ few TeV, whereas the observed weakness of the
Newtonian constant is due to the large size of extra dimensions 
($R\gg $ TeV$^{-1}$). 
The effective Planck scale $M_{P}$ of the 4-dimensional theory is
related to $M_{Pf}$ via the large volume $V\sim R^{N}$ of the internal space
as $M_{P}^{2}\simeq M_{Pf}^{2+N}R^{N}$. Hence, the hierarchy problem between 
$M_{W}$ and $M_{Pf}$ is ''nullified'' but now it can be reformulated as a
question to why the extra dimensions have a size $R$ much larger than
fundamental Planck length $M_{Pf}^{-1}$. At distances larger than the
typical size of these extra dimensions gravitational potential goes to its
standard Newton's form. Moreover, all currently known experimental data
including that of astrophysical as well as cosmological constraints can be
barely satisfied by the above theoretical construction in the case of two or
more extra dimensions, $N\geq 2$ \cite{led}.

Recent developments in string theory have revealed an interesting
possibility that the string scale \ $M_{str}$ \ may be much lower than the
fundamental Planck scale \ $M_{P}$ \cite{Witten} and perhaps as low as $\sim 
$few TeV \cite{AADD, DDG}. As it is well known, any consistent superstring
theory has two parameters: a mass scale \ $M_{str}$ \ and dimensionless
string coupling \ $g_{str}$ \ given by the vacuum expectation value (VEV) of
the dilaton field. Upon the compactification of extra dimensions these
parameters determine the four-dimensional Planck mass \ $M_{P}$ and a single
dimensionless gauge coupling \ $g$ at the string scale $M_{str}$. Thus, in
the context of string theory we have to achieve an unification of the SM
gauge couplings at $M_{str}$, while they are known substantially differ from
each other at currently available energies. 
%(HERE\ I DONT\ UNDERSTAND\ WHY\
%WE\ NEED\ SOME\ NON-ABELIAN\ GAUGE\ SYMMETRY???). 
However, the extrapolation
of gauge couplings from their precisely measured values at $Z$-peak to
higher energies according to the ordinary 4-dimensional renormalization flow
gives the unification scale around $10^{16}$ GeV, 
much higher than the electroweak scale.\footnote{
This is an additional point of the problem of hierarchy of fundamental
scales.}.

One possibility to achieve a low--scale ($O$(TeV)) gauge coupling
unification is to consider the possibility that the number of space-time
dimensions experienced by the SM fields are rised when the SM\ gauge
couplings are running on their way to the unification point, i.e. to assume
the presence of some extra dimensions with radii larger than $M_{str}$. This
leads to change in the evolution of the gauge couplings from 4-dimensional
logarithmic to higher-dimensional power--low and, hence, accelerates
unification in an energy region where the theory becomes high dimensional.
Considerable interest to this possibility was renewed by the first
indication of unification of gauge couplings extrapolated from one--loop
calculation \cite{DDG}. After more accurate test of the minimal scenario it
became evident that in order to justify the unification condition \cite{GR}
an extension of the SM \ or the Minimal Supersymmetric Standard Model (\
MSSM) \ is required. Typically, to improve the unification picture of the
minimal scenario, one considers the models with extra vector--like matter 
\cite{Zura,DQ} or extended SM 
gauge symmetry \cite{PLM} or when the SUSY breaking scale 
is large than the compactification scale \cite{dn}, 
but both are  order of TeV.

In this connection the following question might be naturally raised: 
\textit{What kind of extension of the SM or MSSM 
are needed to achieve the
low-scale gauge coupling unification just in four dimensions without help
of extra dimensions?} If such models can be constructed 
then they can be
also considered from the point of view of a possible resolution of the
unification problem in various higher dimensional models where the SM fields
are restricted to be stuck on a 3-brane and thus do not feel the extra
dimensions at all. 
This might be the situation within the millimeter size
extra dimensions \cite{ADD,AADD} or within the models with non--compact
extra dimensions \cite{RS}. 
%(HERE\ THE\ STATEMENT\ WAS\ NOT\ VERY\ PRECISE:\
%IN\ ADD\ SCENARIO\ THERE\ MIGHT\ BE\ EXTRA\ ON\ THE\ BRANE\ TEV\
%DIMENSIONS, WHILE\ IN\ THE\ RS\ CASE\ THERE\ IS\ ACTUALLY\ NO\ 
%NEED\ TO\ LOCALIZE\ ALL\ FIELDS\ ON\ THE\ BRANE)

In this paper we study systematically the group-theoretical constraints on
the gauge coupling $b$-functions that ensure \ TeV \ scale unification in
the \ SUSY \ as well as in the \ non-SUSY cases. We find a certain set of
new particles which modify properly the running of the gauge couplings
driving them to unify at energies around the \ TeV \ scale. It is remarkable
that such particles can be easily observed at future colluders.

The paper is organized as follows. 
In the next section we present a general
analysis of the gauge coupling unification and determine how the
corresponding $b$-factors should be modified in order to 
have successful low-scale unification. 
Based on this analysis we give some
explicit examples of extra matter multiplets  
which provide the gauge coupling unification at the two-loop level. 
In Section 3 we briefly discuss some
possibilities how to circumvent the proton decay problem 
which we face in low-scale GUT models. 
Finally, in Section 4, we present our conclusions.

\section{Gauge coupling unification at TeV scale }

Let us start by considering general aspects of 
the gauge coupling unification at one-loop level. 
%The present
%experimental data perfectly agree with the fact that for energies up to 
%$\sim 100$ GeV the particle spectrum contains only the SM states.
%
The coupling constants $g_{3,2,1}$ of the standard gauge factors 
$SU(3)\times SU(2)\times U(1)$ at low energies are known with
a quite good accuracy. Namely, the world averages for the values 
$\alpha_{i}=g_{i}^{2}/4\pi$ at $Z$-peak are the following \cite{PDG}: 
\begin{equation}\label{exp}
\alpha_{1}^{-1}(M_{Z})=58.98\pm 0.04,~~~~
\alpha_{2}^{-1}(M_{Z})=29.57\pm 0.03,~~~~
\alpha_{3}(M_{Z})=0.119\pm 0.002.  
\end{equation}
The running constants $\alpha _{i}(\mu)$ at higher energies, 
$\mu >M_{Z}$, can be calculated by the standard renormalization 
group (RG) equations. The fact of the gauge coupling unification 
means that the all three running constants become equal at some 
scale $\mu=M_{U}$, i.e. 
$\alpha _{1}(M_{U})=\alpha_{2}(M_{U})=\alpha _{3}(M_{U})=\alpha _{U}$. 
In principle, starting from some scale $M>M_{Z}$, the theory may include
some extra particle states $F$ in non-trivial representations of 
$SU(3)\times SU(2)\times U(1)$, with masses $M_{F}\geq M$. 
In this case, the one-loop RG equations relating 
$\alpha _{U}$ to $\alpha _{i}(M)$ and $\alpha _{i}(M_Z)$ read as:
\begin{equation} \label{rg}
\alpha _{i}^{-1}(M)=\alpha _{U}^{-1}+\frac{b_{i}^{\mathrm{S}}}{2\pi }
\ln \frac{M_{U}}{M}+\sum_{F}\frac{b_{i}^{\mathrm{F}}}{2\pi }
\ln \frac{M_{U}}{M_{F}}, 
\end{equation}
and
\begin{equation} \label{rgZ}
\alpha^{-1}_{i}(M_Z) = \alpha _{i}^{-1}(M) +  
\frac{b_{i}^{\mathrm{S}}}{2\pi} \ln \frac{M}{M_Z} 
\end{equation}
where  $b_{i}^{\mathrm{S}}$ are the standard one-loop $b$-coefficients.
Namely, in the Standard Model we have 
$b_{i}^{\mathrm{S}}=b_{i}^{\mathrm{SM}}=(41/10,-19/6,-7)$, 
while in the MSSM 
$b_{i}^{\mathrm{S}}=b_{i}^{\mathrm{MSSM}}=(33/5,1,-3)$.\footnote{ 
In the following, for the sake of definiteness, we take 
the effective SUSY scale as $M_Z$.
}
The second term in (\ref{rg}) stands for the contribution of 
extra particles $F$, with one-loop coefficients $b_i^F$ which 
depend on the representation content of the latter. 
%which will be determined below.

%In particular, a multiplet $F(d_3,d_2,Y)$, where $D_3$ and $D_2$ 
%respectively denote the dimensions $SU(3)$ and $SU(2)$ representations 
%and $Y$ is the $U(1)$ hypercharge, gives 
%\begin{equation} \label{deltab}
%b_1^F = \frac{3\xi}{20}Y^2 d_2d_3  , ~~~~  
%b_2^F = \frac{\xi}{3}d_2d_3 C_2(d_2) , ~~~~  
%b_3^F = \frac{\xi}{8}d_2 d_3 C_2(d_3) ,   
%\end{equation}
%where $C_2$ stands for the quadratic Casimir operator 
%for a given representation, and the factor $\xi$ has 
%values  respectively $2/3$, $1/3$ or 1 for $F$ being a chiral fermion, 
%scalar, or chiral supermultiplet of $N=1$ supersymmetry.  

One can introduce the effective coefficients $b_i$ which
extrapolate eqs. (\ref{rg}) as  
\begin{equation}  \label{rg-eff}
\alpha_{i}^{-1}(M)= 
\alpha^{-1}_U + \frac{b_i}{2\pi}\ln \frac{M_U}{M} . 
\end{equation}
>From here follows that 
\begin{equation}  \label{unif1}
\frac{\alpha^{-1}_{i}(M)-\alpha^{-1}_{j}(M)}
{\alpha^{-1}_{j}(M)-\alpha^{-1}_{k}(M)} =
\frac{b_{i} -b_{j} } {b_{j} -b_{k} } ,  ~~~~
i,j,k=1,2,3, 
\end{equation}
and for the difference of the effective $b$-coefficients 
$B_{ij}=b_{i} -b_{j}$ we obtain: 
\begin{equation}  \label{b-eff}
B_{ij} = \frac{B_{ij}^{\mathrm{S}} + 
\sum_F B_{ij}^F} {1 + \frac{1}{2\pi A_{ij}(M) } 
\sum_F B_{ij}^F \ln \frac{M_F}{M} }, 
\end{equation}
where 
$B_{ij}^{\mathrm{S,F}}=b_{i}^{\mathrm{S,F}}-b_{j}^{\mathrm{S,F}}$  
and $A_{ij}(M)=\alpha^{-1}_i(M)-\alpha^{-1}_j(M)$. 
As for the unification scale and the unified gauge constant, 
we have respectively: 
\begin{equation}  \label{MU}
\ln\frac{M_U}{M} = 
\frac{2\pi A_{ij}^{-1}(M)}{B_{ij} } , ~~~~ 
\alpha^{-1}_U =\alpha^{-1}_{1}(M)- A_{12}(M)\frac{b_1}{B_{12}}
\end{equation}

Hence, the criterion for the gauge coupling crossing at one scale 
is encoded into the condition: 
\begin{equation}  \label{unif}
\frac{B_{12}}{B_{23}} = \frac{A_{12}(M)}{A_{23}(M)} 
\equiv R(M),  
\end{equation}
In other words, the theoretical ratio $B\equiv B_{12}/B_{23}$ 
which depends on the extra particle content and their mass splitting, 
should coincide with the quantity $R(M)$ which is 
determined by the experimental values of the gauge coupling 
constants (\ref{exp}). In particular, 
for $M=M_Z$, we have $R(M_Z) = 1.39 \pm 0.03$, with the
uncertainty corresponding to $2\sigma$ error-bars in 
eqs. (\ref{exp}). For $M$ larger than $M_Z$, we obtain 
from (\ref{rgZ}): 
\begin{equation}\label{RM}
R(M)= R(M_Z)\,
\frac{1 -\frac{B_{12}^{\rm S}}{2\pi A_{12}(M_Z)}\ln\frac{M}{M_Z} } 
{1 -\frac{B_{23}^{\rm S}}{2\pi A_{23}(M_Z)}\ln\frac{M}{M_Z} } 
\end{equation}
Therefore, for $M$ in the range up to TeV or so the values 
of $R(M)$ remain very close to $R(M_Z)$ (see Fig. 1). 
In particular, 
in the MSSM we have $R(M)=R(M_Z)$ with a very good accuracy 
while in the SM we obtain 
$R(M) = R(M_Z)[1 - 10^{-2} \ln(M/M_Z)]$, 
where the correction is comparable to uncertainty in 
$R(M_Z)$ itself. 
The values of $R(M)$ for $M > M_Z$ 
calculated in 1 and 2-loops are shown in Fig. 2.

\begin{figure}[t]
%\vspace*{0.7truecm}
\par
\begin{center}
\epsfig{file=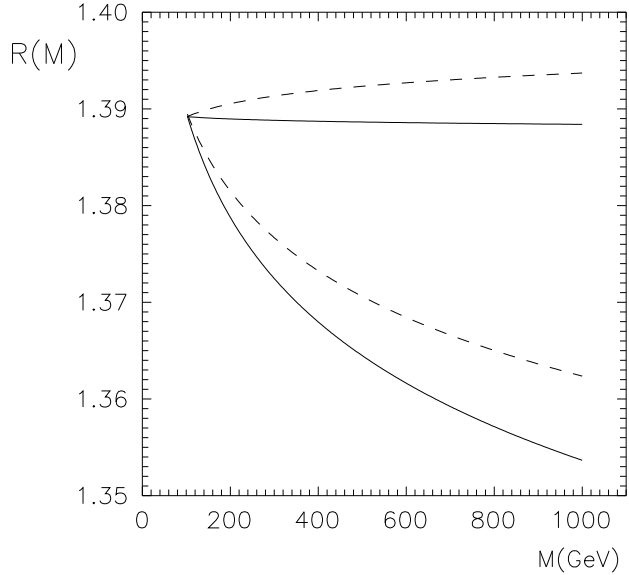,height=8cm}
\end{center}
\caption{{{{{{\protect\small 
$R$ as a function of the intermediate scale $M$ 
for the SM (lower solid) and MSSM (upper solid) at 
one loop. Dash curves correspond to the 2-loop results. 
}}}}}}
\label{fig1}
\end{figure}

In particular, in the SM case, without the extra particle 
contribution, 
the unification condition is not satisfied: we have 
$B=B_{12}^{\mathrm{S}}/B_{23}^{\mathrm{S}}\approx 1.90$, 
about 17 standard deviations away from the range 
$R(M_Z) = 1.39 \pm 0.03$.
%
%$b_{i-j}^{\rm eff} = b_i^{\rm SM}-b_j^{\rm SM}$,   
%$\beta_{12}^{\mathrm{S}}= 109/15$ and $\beta_{23}^{\mathrm{S}}= 23/6$. 
%$\beta_{12}^{\mathrm{S}}= 28/5$ and $\beta_{23}^{\mathrm{S}}= 4$, and so
%
However, in the MSSM we have $B = 1.4$, 
in a wanderful agreement with $R(M_Z)$.
In this way, we have once again demonstrated the remarkable 
success of the supersymmetric grand unification \cite{Goran}.
Since the MSSM yields rather small values of 
$B_{ij}$, namely 
$B_{12}^{\rm S} = 5.6$ and $B_{23}^{\rm S}=4$,   
the gauge coupling unification occurs at very large scale, 
$M_U \simeq 10^{16}$ GeV, which also renders the proton 
to be enough long-living.
%\footnote{
%In the standard model, within the precission of the experimental coupling
%constants at $Z$-peak, $\alpha_1$ and $\alpha_2$ cross at $\mu = 10^{13}$
%GeV, while $\alpha_2$ and $\alpha_3$ cross at $\mu = 3\cdot 10^{16}$
%GeV.}.

On the other hand, lower $M_U$ would need 
the larger values of the coefficients $B_{ij}$ 
which should be achieved due to the contribution of the extra
particles. 
%$B_{ij}\sim 100$.   
In particular, for achieving the unification at the scale 
$M_U$ order few TeV one has to choose the representation content 
of the extra states $F$ so that the coefficients 
$B_{12}$ and $B_{23}$ are large, $O(100)$, and they are 
related as in eq. (\ref{unif}). 
The required correlation $B_{23}=B_{12}/R(M_Z)$  
as well as the values of the unification scale 
as a function of $B_{12}$ is shown in Fig. 2.

\begin{figure}[t]
%\vspace*{0.5truecm}
\par
\begin{center}
\epsfig{file=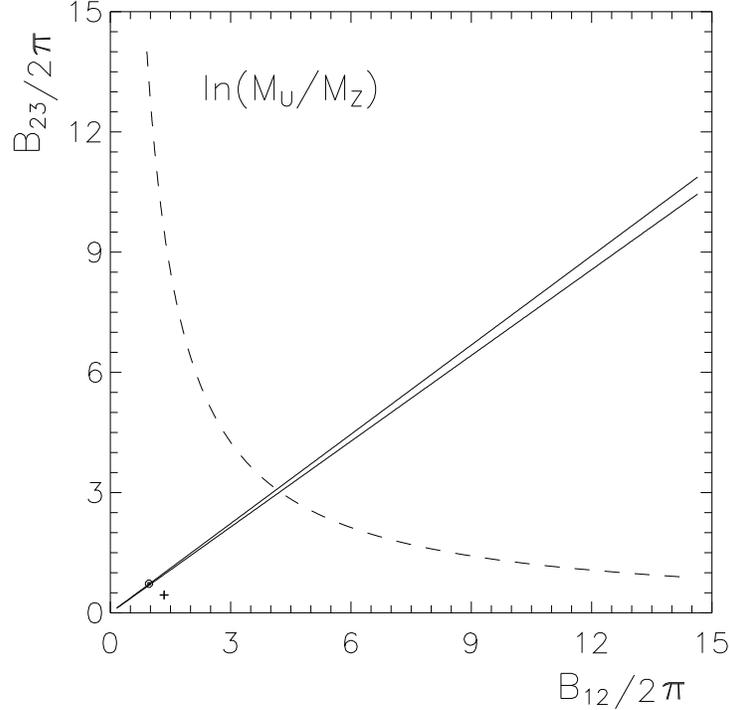,height=10cm}
\end{center}
\caption{{{{{{\protect \small 
Thin area between solid lines indicates 
the correlation between the effective
coefficients $B_{12}$ and $B_{23}$
required by the gauge coupling constant unification 
(\ref{unif}) for the case $M=M_Z$. 
%is taken, with $R_{\rm exp} = 1.39\pm 0.03$. 
The cross and bullet mark the values 
$B_{12}$, $B_{23}$ respectively for the SM and MSSM cases. 
The dashed curve shows the unification scale as a function 
of $B_{12}$. 
}}}}}}
\end{figure}

In order to define better the rules of the game, 
let us assume that extra particles may have masses  
from few hundred GeV to few TeV. 
In the standard model context they can present in the form of 
new scalars or vector-like fermions. In the MSSM context one has 
to introduce chiral superfields in vector-like or self-adjoint 
representations.  
Such massive states would not affect the phenomenology of 
the standard model and also would satisfy the
anomaly cancellation conditions.

Clearly, for achieving the big values of $B_{23}$ in more
economic way, one has to increase $b_2$ without increasing 
much $b_3$. 
In other words, one has to introduce as few as possible extra states 
with the non-trivial colour content, and ultimately one could 
restrict himself only by the colour singlet states. 
Moreover, too many colour states would violate the asymptotic 
freedom of $SU(3)$ so that the gauge coupling crossing could 
occur in the strong coupling regime where the perturbation theory 
is no more valid, or even in the unphysical region 
with the negative $\alpha_U^{-1}$. 
In particular, in order the evolution of $\alpha_3^{-1}$
to higher energies does not change the slope to a negative value, 
we need $\sum_F b_{3}^F$ to be less than 7 in the case of the SM, 
or less than 3 in the MSSM context. 

This simplifies our selection rules for the representation content 
of the extra states. 
The low scale unification of the gauge couplings should be
achieved essentially due to the contributions of the colour singlet 
extra states - sort of heavy leptons. 
We have also to demand that electric charges $Q=T_3 + Y/2$ of  
these states are integer, which means that  
the isospin $T$ and hypercharge $Y$ should be related as  
$Y=2(T+k)$, $k$ being any integer. 

For definiteness, let us consider the supersymmetric case, 
with extra matter in the form of vector-like pairs 
of the colour singlet chiral superfields in $SU(2)\times U(1)$ 
representations $F(D,Y) + \overline{F}(D,-Y)$, 
where $D=2T+1$ is the dimension of $SU(2)$ representation. 
These fields should form massive states with the mass terms 
$M_F \overline{F}F$, $M_F \geq M_Z$.\footnote{The superfields 
in real representations of $SU(2)\times U(1)$, i.e. the ones 
with $Y=0$ and $D$ being the odd number, can have also 
Majorana like mass terms $M_F FF$.}   

Each of the superfields $F$ and $\overline{F}$  
contribute the one-loop $b$ coefficients as follows:\footnote{
In principle, small amount of the colour triplets 
(the heavy exotic quarks) can be also allowed, 
in representations $Q(3,D,Y)+\overline{Q}(\bar3,D,-Y)$, 
with the hypercharges 
$Y=\frac13 + D + 2k$, $k$ being any integer.
Their contributions in the $b$ coefficients are: 
$b_1^Q = \frac{9}{20}DY^2$, $b_2^Q = \frac{1}{4}D(D^2-1)$ 
and $b_3^Q = \frac{1}{2} D$.
In order to maintain the asymptotic freedom of $SU(3)$, 
the $SU(2)$ dimension of these multiplets should not be 
very large, $\sum_Q D_Q < 3$. 
}
\begin{equation}  \label{deltab-l}
b_1^F = \frac{3}{20}DY^2 , ~~~~ 
b_2^F = \frac{1}{12}D(D^2-1), ~~~~
b_3^F = 0 
\end{equation}

Obviously, the existence of extra heavy states, with 
masses in the range from few hundred GeV to TeV
can be directly tested in the future colliders. 
In addition, these states may contain the particles with 
the same quantum numbers as ordinary leptons and quarks, 
which can also have impact for the mass generation 
mechanism of the latter, in the spirit of the mechanism \cite{FN}. 
Since the mass scale of the heavy states is close to the 
electroweak scale, then in general one could expect some 
violation of the universality (unitarity) of the CKM mixing 
and additional contributions to the flavour changing phenomena. 

%\footnote{
%In non-supersymmetric case, the extra states can present 
%in the form of scalars or chiral fermions, in which cases 
%these expressions should be multiplied respectively 
%by the factors $1/3$ or $2/3$. }  
%for $F$ being a scalar, chiral fermion, or chiral supermultiplet
%of $N=1$ supersymmetry.

One can further simplify a situation and consider 
all extra particles having the same mass, $M_F=M$. 
In this case the eq. (\ref{b-eff}) reduces to   
$B_{ij} = B_{ij}^{\mathrm{S}} + \sum_F B_{ij}^F$.  
%B_{ij}^{\mathrm{NS}}$, $B_{ij}^{\mathrm{NS}}\equiv 
Therefore, our goal is to select the representation 
content of the extra $F$ particles so that the non-standard 
contributions in $B_{12}$ and $B_{23}$ are $O(100)$, 
and their ratio is $\approx R(M)$. 
%$B_{12}^{\mathrm{NS}}/B_{23}^{\mathrm{NS}} 
In particular, one can choose the representations 
which predict $B=1.4$, as it holds in the case of the MSSM. 

Clearly, many different solutions can be envisaged. 
%At this point one can consider different strategies. 
For example, one can consider extra states consisting 
exclusively 
of heavy replicas of the ordinary lepton species, 
$L(2,-1)+\overline{L}(2,1)$ and 
$\overline{E}(1,2) + E(1,-2)$,  
for simplicity all located at the same scale $M$. 
Then, if we take their multiplicities as $N_L=2N$ and $N_E=3N$, 
with $N$ being any integer number, 
we obtain $b_{1}^{\rm NS}=4.8N$ and $b_{2}^{\rm NS}=2N$, 
and so $B^{NS}=1.4$. 
On the other hand, for having the low scale unification, 
$N$ should be large number. 
For example, for $M=400$ GeV, the unification
scale $M_{U}\simeq 10$ TeV can be obtained only if 
$n=18$.\footnote{
Alternatively, one could take the same number of these species, 
$N_L=N_E$ but consider the mass splitting between these states. 
For instance, $M_U\simeq 10$ TeV can be achieved by taking 
42 copies for each of $L+ \overline{L}$ and $\overline{E}+E$, 
with $M_E=200$ GeV and $M_L=1$ TeV. 
%Certainly, neither case looks very beautiful.
}

% which repeat the quantum numbers of the ordinary leptons, 

Another possibility is to seek for just one big representation 
which could do a job alone. 
A candidate we have found is a vector-like multiplet 
$F(7,8)+\overline{F}(7,-8)$. 
It leads to $b_1^{F} = 134.4$ and $b_2^{F} = 56$, 
and hence $B^{\rm NS}=1.4$. These coefficients are enough big, 
$B_{12}=78.4$, to provide $M_U$ order TeV.
And finally, one can consider a mixed situation, containing some big
representations appended by few small ones. Some possible candidates are
given below.

A subtle questions which arise here is whether the 
solutions found in 1-loop approximation will be stable against 
2-loop corrections. This question seems most challenging for the 
solutions with large representations in which case the 2-loop 
RG effects are expected to be significant.
Therefore, one needs to examine the 2-loop RG equations: 
\begin{equation}
\frac{d\alpha _{i}}{dt}=-\frac{b_{i}}{2\pi }\alpha _{i}^{2}-\frac{1}{8\pi
^{2}}\alpha _{i}^{2}\sum_{j=1}^{3}b_{ij}\alpha _{j}
\end{equation}
where $b_i =b_i^{\mathrm{S}}+\sum_{F}b_{i}^{F}$
are the 1-loop $b$-coefficients, with $b_i^F$ fiven as in 
eq. (\ref{deltab-l}), and 
$b_{ij} =b_{ij}^{\mathrm{S}}+\sum_{F}b_{ij}^{F}$
are the 2-loop ones. In the MSSM the standard 2-loop coefficients 
are given by the matrix: 
\begin{equation}\label{matriz}
b_{ij}^{\mathrm{S}}=\left( 
\begin{array}{ccc}
199/25 & 27/5 & 88/5 \\ 
9/5 & 25 & 24 \\ 
11/5 & 9 & 14
\end{array}
\right) ,  
\end{equation}
while the non-standard contributions can be calculated as 
\cite{Jones}: 
\begin{eqnarray}\label{2loop}
&& 
b_{11}^{F} =\frac{9}{100}DY^{4}, ~~~~  
b_{12}^F = 3b_{21}^F = \frac{3}{20}D(D^{2}-1)Y^{2} , 
\nonumber \\
&& 
b_{22}^F =\frac{1}{12}(D^{2}+3) D(D^{2}-1) ,
%b_{21}(L) =\frac{B}{80}d(d^{2}-1)Y^{2}  
\end{eqnarray}
%
%where for SUSY case A = 2, B = 4, %C = 6, K = 3 
%and for the non-SUSY case A=10/3, B= 2. %  C=34/3 \,  D=2,  K = 11/3.

The two loop corrections significantly affect the solutions 
with small representations.  
%One would expect that the situation would become 
%worse for the solutions with large representations 
However, a remarkable thing occurs for the solutions with 
large representations:   
the fact of the gauge coupling unification remains robust, 
whereas naively one would expect strong violations since the 
2-loop effects normally become bigger with bigger representations. 
%(As for the unification scale, it
%changes strongly with the 2-loop corrections.)

The reason of the stability of the unification condition 
has a simple explanation. To
demonstrate this, it is convenient to approximate the two-loop RG
predictions as follows: 
\begin{equation}  \label{rg-2}
\alpha_{i}^{-1}(M)= \alpha^{-1}_U + 
\frac{b_i}{2\pi}\ln \frac{M_U}{M} + \frac{1}{4\pi} \sum_{j=1}^3 
\frac{b_{ij}}{b_{j}} \ln \frac{\alpha_U}{\alpha_j(M)}
\end{equation}
One can rewrite these equations in a form analogous to 
1-loop extrapolation (\ref{rg-eff}):  
\begin{equation}
\alpha_{i}^{-1}(M)= 
\alpha^{-1}_U + \frac{b_i^{(2)}}{2\pi}\ln \frac{M_U}{M}
\end{equation}
where 
\begin{equation}
b_i^{(2)} = b_i + \frac12 \sum_{j=1}^3 
\frac{b_{ij}}{b_j} X_j , ~~~~ 
X_j = \frac{\ln(\alpha_U/\alpha_j(M))}{\ln(M_U/M)} . 
\end{equation}
Then the gauge coupling crossing condition becomes 
\begin{equation}  \label{unif2}
B^{(2)} \equiv \frac{B_{12}^{(2)}}{B_{23}^{(2)}} = R(M) . 
\end{equation}
Hence, the stability of the 1-loop solution 
with satisfying (\ref{unif}) implies that 
the ratio of 2-loop factors 
$B^{(2)} \equiv B^{(2)}_{12}/B^{(2)}_{23}$
should remain close to the ratio of 1-loop factors 
$B=B_{12}/B_{23}$.

This is what actually happens when the RG evolution of 
couplings is dominated by one big representation $F(D,Y)$, 
due to a conspiracy between the 1- and
2-loop coefficients (\ref{deltab-l}) and (\ref{2loop}). 
Indeed, one obtains:
\begin{equation}
B_{12}^{(2)F} = 
B_{12}^F \left[1+ \frac{2}{D}(X_1 + 3X_2)\right] - 2X_2 , ~~~~ 
B_{23}^{(2)F} =
B_{23}^F \left[1+ \frac{2}{D}(X_1 + 3X_2)\right] + 2X_2
\end{equation}
and so  
$B^{(2)} \simeq B$, 
%$B^{(2)}_{12}/B^{(2)}_{23}\simeq B_{12}/B_{23}$, 
with about a per cent correction caused by
the terms $2X_2$.

We found the solutions 
by solving numerically the precise 2-loop equations as well. 
%As input parameters we used the experimental data given in (\ref{exp}).
%Extra matter was assumed to have a common mass $M$. 
Namely, first we justify the unification condition (\ref{unif}), 
by selecting the particle states with appropriate quantum numbers 
which could ensure the desired modification of $b$-factors 
at one-loop. Then we test these 
solutions by numerical analysis at two-loop level.\footnote{ 
Higher-loop corrections are expected to be negligible as we have 
a logarithmic running of gauge couplings and all gauge
couplings are in perturbative region.
}
%where we use standard threshold corrections and also included
%conversion from the $\overline{MS}$ scheme to the 
%$\overline{DR}$ one. 

As it was expected from the above analysis, the case of large 
multiplet $F(7,8)+\overline{F}(7,-8)$ 
which satisfies the 1-loop criterion of the gauge coupling 
crossing (\ref{unif}), $B=1.4$, is perfectly stable 
in two-loops. However, the value of unification scale 
changes significantly. Namely, by taking the mass scale 
$M=400$ GeV, we had $M_U^{(1)} = 3.3$ TeV at 1-loop, 
while at 2-loops we obtain $M_U^{(2)} = 2$ TeV. 

The list of some mixed solutions is given below.  They all imply $B=1.4$
at 1-loop and are stable against 2-loop RG analysis. We show the
$SU(2)\times U(1)$ content and multiplicities of the chiral superfields
(plus their conjugates) and the unification scales at 1- and 2-loops
respectively.  
\begin{eqnarray}
(6,7)~+~3\times (2,1)~+~(1,2) &:&M_{U}^{(1)}=8.2~\mathrm{TeV}
,~~~M_{U}^{(2)}=4.8~\mathrm{TeV}  \nonumber   \\
2\times (5,6)~+~2\times (3,2) &:&M_{U}^{(1)}=4.6~\mathrm{TeV}
,~~~M_{U}^{(2)}=3.3~\mathrm{TeV}  \nonumber \\
2\times (4,5)~+2\times (3,2) &:&M_{U}^{(1)}=21~\mathrm{TeV}
,~~~M_{U}^{(2)}=15~\mathrm{TeV} \label{susyc} 
\end{eqnarray}

In the case of solutions with only small representations 
$L+\overline{L}$ and $E+ \overline{E}$, with $N_L=36$ and 
$N_E=54$, the coupling 
crossing does not occur anymore within $2\sigma$ error-bars in
$\alpha_3(M_Z)$, and it requires about $4.5\sigma$ deviation. 
In order to obtain the coupling crossing at 
$2\sigma$ level, on has to change the number of fields, 
namely to take $N_L=35$ and $N_E=54$,  
%Namely, the case $35\times (2,1) + 54\times (1,2)$ should be taken, 
which at one-loop would correspond to $B=1.45$. 
On the other hand, the unification scale does not change
substantially and remains around $10$ TeV.

\section{GUT picture and proton stability}

An interesting issue is whether the coupling constant crossing 
can really correspond to the possibility that at the scale $M_{U}$ 
of a few TeV three gauge gauge factors 
$SU(3)\times SU(2)\times U(1)$ are indeed embedded
in some grand unified group. 
The main obstacle for the realization of the
consistent GUT scenario is the problem of proton stability. 
Indeed, if the quark and lepton fields are unified into the $SU(5)$
multiplets $\bar{5}_{k}$ and $10_{k}$, $k=1,2,3$, then the processes 
mediated by the heavy gauge bosons $X$ and $Y$ of $SU(5)$ with masses 
$\sim M_{U}$ will lead to the catastrophically fast proton decay. 
%(THE\ RATE\ ACTUALLY\ 10**4 TIMES
%SMALLER THAN\ THE\ WEAK\ DECAY\ RATES).

Nevertheless, several ways out can be envisaged. 
One possible scenario for the suppression of the proton decay can be thought 
as follows. 
Suppose the gauge $SU(5)$ theory which is in the confining phase outside the
3+1-dimensional domain wall (3-brane) produced by some master 
field $\phi$, a singlet of $SU(5)$.  
However on the 3-brane, due to the appropriate coupling with the master
field $\phi$ , 
some adjoint scalar field $\Sigma$ is triggered to 
develop the  $SU(3)\times SU(2)\times U(1)$ preserving VEV, 
$\langle\Sigma\rangle = V \cdot {\rm diag}(2,2,2,-3,-3)$.   
Then, according to the Dvali-Shifman mechanism \cite{p3}, all
gauge fields of the standard model are localized on the 3-brane, 
while the massive (on the 3-brane) $X$ and $Y$ bosons freely propagate 
in the bulk. Now, assuming also that chiral matter fields in $\bar{5}+10$
representations of $SU(5)$ are confined on the 3-brane 
%(WHAT\ CHIRALITY\ PROBLEM\ WE\ WILL\ HAVE\ HERE?), 
their effective 4-dimensional couplings to $X$ and $Y$ bosons could
be suppressed by the volume-factor comming from the integration out of 
extra coordinates in the effective Lagrangian: 
\begin{equation}
g_{X,Y}^{(4-{dim})}=\frac{1}{\sqrt{V_{N}}} g_{SU(5)}^{(bulk)},
\label{pd3}
\end{equation}
where $V_{N}\sim R^{N}$ is the volume of compact internal space with 
$N$ extra dimensions of radius $R$. 
The rough estimation shows that the case of 
$N=2$ extra dimensions with $R\sim 10^{-3}$ mm is perfectly safe 
to satisfy experimental bounds on the proton lifetime even
if the unification scale is around TeV.

The obvious trouble with the above mechanism is related with the colour
triplet partner of the electroweak doublet. The point is that, as long as
gluons are confined on the brane the coloured Higgs has to be restricted on
the brane as well, due to the colour flux conservation argument \cite{p3}.
This coloured Higgs (as well as its superpartner, Higgsino) can mediate fast
proton decay, unless it is completely decoupled from the quarks and leptons.
Such a decoupling can be indeed achieved in some extended GUTs (say, in 
$SO(10)$ GUT) thanks to the special pattern of the GUT symmetry breaking 
(for more details see \cite{p4}). 
Another way is to simply remove the fundamental
Higgs from the theory, attributing the electroweak symmetry breaking to some
dynamical mechanism involving say the top-antitop condensation \cite{p5}.

Another possibility of keeping baryon and /or lepton numbers approximately
conserved in 4-dimensions and thus suppressing the proton decay up to a
desired level is offered by the mechanism of ref. \cite{p2}. It reveals to
the field-theoretic localization of chiral matter on a fat 3+1-dimensional
domain wall (3-brane) when the quarks and leptons are localized at a
different points along the extra coordinate. This mechanism is also
compatible with GUT models, providing that GUT symmetry, say $SU(5)$, is
broken down to the $SU(3)\times SU(2)\times U(1)$ by the VEV 
%$<\Sigma >$ $%= v \, diag(2,2,2,-3,-3)$ 
of the bulk adjoint field $\Sigma $. The
higher-dimensional model is initially vector-like, so along with the usual 
$(\bar{5}+10)$ representations for each family of ordinary quarks and
leptons there are mirror fermions $(5+\overline{10})$  as well. 
In the bulk
the Dirac masses of the quarks and leptons residing in quintuplets and
decuplets are splitted as a result of GUT symmetry breaking according to the
following equations: 
\begin{equation}
\bar{5}\left( \langle\Sigma\rangle + M_{5}\right) 5, ~~~~
\overline{10}\left(\langle\Sigma\rangle + M_{10}\right) 10.  
\label{pd1}
\end{equation}
Here we omit Yukawa constants and family indeces for simplicitly; $M_{5}$
and $M_{10}$ are $SU(5)$ invariant masses for the quintuplets and decuplets,
respectively.

The matter fields above are assumed to couple with the master scalar field 
($SU(5)$-singlet) $\phi $ as well. This scalar field ''produce'' the domain
wall $\phi (x_{5})=\phi _{0}\tanh (\mu x_{5})$ with a certain thikness 
$\mu^{-1}$. 
Then, as it is well known, only the chiral matter (e.g. $(\bar{5}+10)$) 
get localized on the domain wall. The points where the wave
functions of the quarks and leptons are peaked are actually different
(schematically we denote them as $x_{5}^{q}$ and $x_{5}^{l},$ respectively)
due to the precence of the $SU(5)$-breaking VEV of $\Sigma$ in (\ref{pd1})
and are determined by the equations: 
\begin{equation}
\overline{5}\left(\langle\Sigma\rangle + M_{5}+ \phi (x_{5})\right) 5=0,
~~~~~
\overline{10}\left(\langle\Sigma\rangle + M_{10}+\phi (x_{5})\right) 10=0.  
\label{pd2}
\end{equation}
Thus, any transition between quarks and leptons in 4 dimensions will contain
an exponetial suppression factor 
$\exp \left( -\mu^{2}(x_{5}^{q}-x_{5}^{l})^{2}\right)$, 
since the interactions are non-local
in extra dimension. This exponential factor suitably adjusted could indeed
suppress the proton decay up to a desired level.

Perhaps the most natural
realization of such GUT models can be found within the higher-dimensional
theories when certain symmetries inherited from the compactified internal
space can be used to prevent the rapid proton decay. One of such
intrinsically higher-dimensional mechanism has been proposed in \cite{DDG}
where the minimal SUSY $SU(5)$ was considered in 5 dimensions compactified
on $S^{1}/Z_{2}$ orbifold. The chiral matter (i.e. ordinary quarks and
leptons and their superparners) in \cite{DDG} is assumed to be located at
the orbifold fixed point, while gauge fields and possibly Higgs fields as
well are allowed to propagate in the full 5-dimensional bulk. Then, assuming
that $X$ and $Y$ bosons (and the colored Higgs as well) are
odd under the $Z_{2}$ orbifold parity, one can totally decouple them from
the quarks and leptons, so that they could not be responsible for the proton
decay \footnote{
The minimal proposal of ref. \cite{DDG} do not seem fully satisfactory since
the scalar leptoquarks remained couple with quarks and leptons could also
mediate unacceptable rapid proton decay. For further developments of the
mechanism of ref. \cite{DDG} see \cite{p1}.}.

Finally, one could construct the GUT models with absolutely stabile proton 
just in 4 dimensions. 
For some models constructed in the past see e.g. \cite{p6}.
The obvious candidates for such a GUTs are the models with discretely
covered GUT symmetry, such as $\left( SU(3)\right) ^{3}$ trinification or 
$SU(N)\otimes SU(N)$-type theories with approprately chosen matter
representations. In such models typically there are no $X$ and $Y$ gauge
bosons (or they are not responcible for the transition of ordinary quarks
into leptons and vice versa) and thus there is no gauge mediation of the
proton decay. As to the coloured Higgs (Higgsino) mediation of proton decay
one can still use one of the mechanism described above. 

Moreover, one can
construct the GUT models based on the simple groups as well. As an
instructive example we scotch here $SU(5)$ GUT with special fermion
assignment. Namely, imagine that the theory contains 5 states $\bar{F}
_{1,2}\sim \bar{5}$ and $T_{1,2,3}\sim 10$, and 3 anti-states $F$ and $\bar{T
}_{1,2}$ per each generation, among the other possible vector-like states.
Imagine also that $SU(5)$ symmetry is broken by the fields $\Phi $ and $
\Omega $, both in the reducible $24+1$ representations, having the following
''orthogonal'' VEV patterns: $\langle \Phi \rangle =V_{1}\cdot \mathrm{diag}
(1,1,1,0,0)$ and $\langle \Omega \rangle =V_{2}\cdot \mathrm{diag}(0,0,0,1,1)
$. There is also one extra $24+1$ field $\Sigma $ which develops VEV of the
form $\langle \Sigma \rangle =\sigma \cdot \mathrm{diag}(1,1,1,-1,-1)$.
Note, however, that all this fields and their VEVs more naturally look in
the context of GUTs higher than $SU(5)$. E.g. in 
$SU(5)\otimes SU(5)$ they can emerge from the mixed representations 
like $(5,\bar5)$. Let us now consider
the following superpotential terms: 
\begin{equation}
\label{su5}
\left( \bar{F}_{1}F+T_{1}\bar{T}_{1}\right) \Phi +\left( \bar{F}_{2}F+T_{2}
\bar{T}_{1}\right) \Omega +T_{3}\bar{T}_{2}\Sigma  
\end{equation}
Then, the VEVs of $\Phi ,$ $\Omega $ and $\Sigma $ pick up the $L\subset $ $
\bar{F}_{1}$, $d^{c}\subset $ $\bar{F}_{2}$, $e^{c}\subset $ $T_{1}$, 
$u^{c}\subset $ $T_{2}$ and $Q\subset T_{3}$ as an ordinary massless
quarks and leptons of the SM, while all other states get masses from the
couplings in (\ref{su5}). At this stage, keeping only massless states as
an external ones and while considering all massive states as those of
possible intermediate, one can define five separately conserved global
charges. They are: 
\begin{eqnarray}
C(\bar{F}_{1}) &=&N(e)+N(\nu )  \nonumber \\
C(\bar{F}_{2}) &=&N(d^{c})  \nonumber \\
C(T_{1}) &=&N(e^{c})  \nonumber \\
C(T_{2}) &=&N(u^{c})  \nonumber \\
C(T_{3}) &=&N(u)+N(d)  \label{ch}
\end{eqnarray}
where $N$ denotes the particle number operator. Now, whatever mechanism is
responsible for the generation of the masses for the SM quarks and leptons
above, it is evident that $\bar{F}_{1}$ and $\bar{F}_{2}$ will couple to $
T_{1}$ and $T_{3},$ respectively and $T_{2}$ will couple to $T_{3}$ as well.
Thus, the only two combinations of charges in (\ref{ch}) will survive as an
unbroken ones. Namely, 
\begin{eqnarray}
Q_{1} &=&N(e)+N(\nu )+N(e^{c})\equiv L  \nonumber \\
Q_{2} &=&N(u)+N(d)+N(u^{c})+N(d^{c})\equiv B  \label{ch1}
\end{eqnarray}
separately conserved. They are the lepton $L$ number and the
baryon number $B$. 
Thus, the proton is stabile in all orders of perturbation
theory. 
%(DO\ WE\ REALY NEED\ TO\ CONSIDER SU(5)*SU(5) model?).

The following remark is in order. 
One could ask, what can be the origin of the "leptonic" 
fragments $F+\overline{F}$ in big representations of 
$SU(2)\times U(1)$ considered in previous section,  
which are needed to properly correct the RG evolution of the 
gauge couplings. Clearly, these can naturally emerge 
from the GUT superfields in big representations. 
For example, consider the $SU(5)$ superfields in 
$2$-index symmetric representations $T_{ab}\sim 15$ and 
$\overline{T}^{ab}\sim \overline{15}$.  
One can consider the couplings $T\Phi\overline{T}$ 
with the Higgs $\Phi$ having the VEV 
$\propto {\rm diag}(1,1,1,0,0)$. 
Clearly, this would give order $M_U$ mass to all fragments 
in $T$ and $\overline{T}$ apart 
of the fragments $F(3,2)+\overline{F}(3,-2)$ 
(all indices from $SU(2)$ subgroup).  
Any other "leptonic" superfield in representations 
$F(D,Y)+\overline{F}(D,-Y)$ can be left light by the 
interaction with the Higgs $\Phi$ 
%in view of such missing VEV mechanism, 
by proper choice of the 
corresponding big $SU(5)$ representation, while any fragment 
with at least one colour index would get mass order $M_U$. 
Thus, the extra particle states needed for correcting 
the RG evolution of the gauge couplings for achieving 
the TeV scale unification can be obtained by the same 
"missing VEV" mechanism which also guarantees the 
fermion mass arangement rendering the proton stable 
and also leaves the Higgs doublets light.

\section{Conclusions}

We have demonstrated that the 
the unification of $SU(3)\times SU(2) \times U(1)$
gauge couplings at very low scale 
can be achieved in four dimensions with the help of 
not so little friends - some extra particles in rather 
big representations of $SU(2)\times U(1)$ at some 
intermediate scales between few hundred GeV and TeV. 
In this case no power like RG evolution has to be invoked  
in extra dimensions. 
These particles can be directly observed at the future 
accelerators and have many phenomenological implications. 
Theoretical models can be constructed in which the 
garnd unification occurs at TeV scales but proton remains 
stable.  

\vspace{5mm}
\noindent {\bf Acknowledgments.}
\vspace{3mm}

We gratefully acknowledge helpful discussions with Gia Dvali, Gregory
Gabadadze, and especially with Goran Senjanovi\'c who was extremely
enthusiastic about the TeV scale grand unification. 
The work of Z.B. was partially supported by the MURST research grant
"Astroparticle Physics" and that of A.B.K. by the Academy of Finland 
under the Project No. 163394.

\end{document}